\documentclass[aps,prl,
twocolumn, 
amsfonts,amssymb,amsmath,
tightenlines,
twoside,
letterpaper,
bibnotes,
]{revtex4}

\usepackage{bm,color,mathrsfs,graphicx}

\begin{document}

\title{Electromagnon Resonances in the Optical Response of Spiral Magnets}

\author{A. Cano}
\email{cano@esrf.fr}
\affiliation{
European Synchrotron Radiation Facility, 6 rue Jules Horowitz, BP 220, 38043 Grenoble, France}

\date{\today}

\begin{abstract}
The optical response of spiral magnets is studied, with special attention to its electromagnon features. 
\textcolor{black}{
We show that these features trace back to the resonant magnetoelectric response resulting from the spiral ordering (irrespective of any concomitant ferroelectricity). This response, being magnetoelectric in nature, not always can be reduced to an effective electric permittivity.}
We argue that electromagnons in spiral magnets can produce, in addition to the observed peaks in the optical absorption of multiferroics, a (dynamically enhanced) optical rotation and a negative refractive index behavior.
\end{abstract}


\maketitle

The strong interplay between magnetism and ferroelectricity observed in a new generation of ferroelectromagnets (or multiferroics) has prompted a renewed interest on magnetoelectric (ME) phenomena. In TbMnO$_3$, for example, the electric polarization can be flopped by applying a magnetic field \cite{Kimura03} and, conversely, the chirality of its magnetic structure can be changed by applying an electric field \cite{Yamasaki07}. The dynamic counterpart of these cross-coupling effects is the existence of hybrid magnon-polar modes, i.e., the so-called electromagnons, which also have been observed in the form of absorption peaks in optical experiments \cite{Pimenov06}. The relatively large magnitude of these ME effects makes this type of materials very attractive as novel memory elements, optical switches, etc.

The ME response is known to be an important ingredient in the electrodynamics of conventional magnetoelectrics \cite{O'Dell,Landau-Lifshitz_ECM,Arima08,Krichevtsov93}. 
In Cr$_2$O$_3$, for example, this response alone suffices to produce a nonreciprocal optical rotation \cite{Krichevtsov93}. In this paper we provide a continuum medium description of the dynamical ME effect in spiral magnets. 
\textcolor{black}{
We show that, in contrast to the static case, the dynamic ME effect has to be described in terms of two ME response tensors. The reason lies in the different reaction to the external perturbations: the electric field is seen as a force, while the magnetic field as a torque. 
Moreover, the frequency dispersion of these ME tensors reveals both magnon and polar modes in the form of resonances. These resonances are further shown to be responsible for electromagnon features, as observed in the aforementioned optical experiments on multiferroics (see also \cite{Sushkov07,Sushkov08,Takahashi08,Kida08}). We also show that these features, being genuinely ME in origin, not always can be reduced to an effective electric permittivity as in previous interpretations.}
We also discuss briefly the possibility of having a dynamically enhanced nonreciprocal optical rotation and a negative refractive index behavior due to such a genuine ME response. 

Let us begin by recalling that, at the static level, the most general linear response of a (homogeneous) medium to the electric and magnetic fields, $\mathbf E$ and $\mathbf H$ respectively, can be expressed by the constitutive relations \cite{O'Dell,Landau-Lifshitz_ECM}:
\begin{subequations}\begin{align}
\mathbf P &=\hat \chi^{e}\mathbf E + \hat \alpha \mathbf H,\label{em_coupling}
\\
\mathbf M &=\hat \alpha ^\text{T}\mathbf E + \hat \chi^{m}\mathbf H.\label{me_coupling}
\end{align}\label{constitutive}\end{subequations}
Here $\mathbf P$ and $\mathbf M$ represent the electric and magnetic polarization, $\hat \chi^{e}$ and $\hat \chi^{m}$ are the electric and magnetic susceptibilities, and $\hat \alpha$ is the ME tensor of the medium. The same tensor $\hat \alpha$ enters in these two equations ($\alpha_{ij}^\text{T} =\alpha_{ji}$) because it traces back to the same coupling $-\alpha_{ij}E_iH_j$ in the free energy of the system. Only a restricted number of magnetic symmetry classes allow for this linear coupling, in which case the system is termed as magnetoelectric. 

In the case of spiral magnets the inhomogeneous ME effect \cite{Baryakhtar83,Cano08,Mills08} is always at work. This effect describes the (universal) coupling between the electric polarization and nonuniform distributions of magnetization. For our purposes it can be taken as \cite{note}
\begin{align}
-f\mathbf P \cdot [\mathbf M ( \nabla \cdot \mathbf M) - (\mathbf M \cdot \nabla )\mathbf M].
\label{F_me}\end{align}
\textcolor{black}{
This coupling is believed to be behind the ferroelectricity of $R$MnO$_3$ compounds, as comes from the observation that a cycloidal magnetization generates a term $-\mathbf P \cdot \mathbf E_\text{eff}$ out of this coupling, where the constant $\mathbf E_\text{eff}$ reflects the lack of inversion symmetry in the cycloidal. 
This is the so-called spiral or spin-current mechanism for ferroelectricity, in which the polarization appears in the plane of the cycloidal perpendicular to its wavevector \cite{Katsura05}. To describe more complicated cases 
one has to go beyond \eqref{F_me} and consider different sublattice magnetization fields and/or the transformation properties of $\mathbf P $ and $\mathbf M$ under the elements of the corresponding point group \cite{Mostovoy08}.}

To extend Eqs. \eqref{constitutive} to the frequency domain we have to deal with the dynamics of the system. In this dynamics the coupling \eqref{F_me} 
\textcolor{black}{
already leads to} the hybridization of magnons with polar modes (see below). 
\textcolor{black}{
It is likely that a more elaborated version of this coupling is needed to explain, e.g., the electromagnon selection rules observed in multiferroics for which the multisublattice magnetic structure seems to play a role \cite{Sushkov08}. Nevertheless we restrict ourselves to the ``minimal'' coupling \eqref{F_me} because, already at this level, the dynamical ME effect can be shown to be richer than noticed before. 
Compared to the static case, for instance, the dynamical ME effect exhibits the following asymmetry. Any hybridization implies that magnons can act as effective (time-dependent) electric fields for phonons and {\it vice versa}, phonons can act as effective magnetic fields for magnons. Then charges will try to follow the mangon-induced field that is pushing them back and forth, whereas spins will tend to precess about the phonon-induced one (here we have a torque). This eventually translates into two dynamical ME tensors, in contrast to the static case \eqref{constitutive}.}

Let us compute these dynamical tensors. In the presence of an electromagnetic radiation, electric and magnetic polarizations will deviate with respect to the corresponding background distributions: $\mathbf P = \mathbf P ^{(0)} + \mathbf p$ and $\mathbf M = \mathbf M ^{(0)} + \mathbf m$.
These deviations $\mathbf p$ and $\mathbf m$ are assumed to be small (proportional to the external fields), so the equations of motion for $\mathbf P$ and $\mathbf M$ can be linearized with respect to these quantities. For the sake of simplicity, we restrict ourselves to background magnetizations containing only one periodicity (i.e., with wavevectors $\pm \mathbf Q$). Thus, if we take the equation of motion for the electric polarization, in Fourier space we get
\begin{widetext}
\begin{align}
&\hat A(\mathbf q,\omega) \mathbf p (\mathbf q,\omega) 
\approx 
\mathbf E(\mathbf q,\omega) 
+ 2i f \sum _{\mathbf q' =\pm \mathbf Q } 
\big[\big(\mathbf q' \cdot \mathbf M^{(0)}(\mathbf q')\big) \mathbf m(\mathbf q -\mathbf q',\omega) 
- 
\mathbf M^{(0)}(\mathbf q')\big(\mathbf q' \cdot \mathbf m(\mathbf q - \mathbf q',\omega)\big) 
\big]
\label{motion_p}\end{align}
\end{widetext}
in the limit $q \ll Q$. 
Here $\hat A$ represents the inverse electric susceptibility (in the absence of ME coupling $\hat A^{-1} \equiv \hat \chi^{e} $). In this equation we can see that, in fact, polar modes are linearly coupled with the deviations $\mathbf m (\mathbf q \pm \mathbf Q, \omega)$ of the magnetic structure by virtue of the modulation of this latter, i.e., we have electromagnons. Close to the 
\textcolor{black}{magnon frequencies, as in ordinary antiferromagnets,} the dynamics of the magnetization is expected to be described by the Landau-Lifshitz equation. Then, the nonlinear character of this equation, together with the non-uniform magnetic background $\mathbf M ^{(0)}$, make it possible the linear coupling between these excitations and long-wavelength external fields (see e.g. \cite{Cano08,Belitz06}). 
\textcolor{black}{
Neglecting the ME coupling here for a while we have}
\begin{align}
\mathbf m (\mathbf q \pm \mathbf Q,\omega ) &\underset{q\ll Q}{=} \hat \chi^{m,\pm \mathbf Q}(\mathbf q,\omega) \mathbf H(\mathbf q,\omega), 
\label{chi_m}
\end{align}
where the poles of $\chi^{m,\pm \mathbf Q}$ are associated with the characteristic excitations of the modulated structure \cite{Cano08,Belitz06}. Substituting this expression into \eqref{motion_p} we obtain
\begin{align}
\mathbf p (\mathbf q,\omega) 
&= \hat \chi^{e} (\mathbf q,\omega) 
\mathbf E (\mathbf q,\omega) 
+ \hat \alpha (\mathbf q,\omega) \mathbf H(\mathbf q,\omega), 
\label{d-em_coupling}
\end{align}
where 
\begin{widetext}
\begin{align}
\alpha_{ij} (\mathbf q,\omega) = 
2i f \sum_{\mathbf q' = \pm \mathbf Q}
q_k'M_{k'}^{(0)}(\mathbf q')  
(
\delta_{i'j'} \delta_{kk'} 
- 
\delta_{i'k'} \delta_{kj'} )
\chi_{ii'}^{e}(\mathbf q,\omega)\chi^{m,-\mathbf q'}_{j'j}(\mathbf q,\omega).
\label{d-em-tensor}
\end{align}
\end{widetext}
The constitutive equation \eqref{d-em_coupling} replaces \eqref{em_coupling} for dynamical processes. 
\textcolor{black}{
The dynamical ME tensor $\hat \alpha$ traces back to the magnon-phonon hybridization resulting from the ME coupling [Eq. \eqref{F_me} in our case].
Both these modes yield resonances in $\hat \alpha$, and these resonances are further responsible for electromagnon features in optical experiments (see below).}

As mentioned before, the fact that polarization and magnetization dynamics are different produces a certain asymmetry in the dynamical ME response. 
Carrying out similar manipulations the equation of motion for the magnetization can be reduced to an expression analogous to \eqref{me_coupling}:
\begin{align}
\mathbf m(\mathbf q,\omega) 
&=\hat \beta(\mathbf q,\omega) \mathbf E(\mathbf q,\omega) + \hat \chi^{m} (\mathbf q,\omega) 
\mathbf H(\mathbf q,\omega) , 
\label{d-me_coupling}
\end{align}
where 
\begin{widetext}
\begin{align}
\beta_{ij}(\mathbf q,\omega) = i \gamma f \epsilon_{ii'i''} 
\sum_{\mathbf q' = \pm \mathbf Q} q'_k 
M_{k'}^{(0)}(\mathbf q')M_{i''}^{(0)}(-\mathbf q')
(
\delta_{j'i'} \delta_{kk'} 
- 
\delta_{j'k'} \delta_{ki'} )
\chi_{j'j}^{e}(\mathbf q,\omega),
\label{d-me-tensor}
\end{align}
\end{widetext}
with $\gamma$ the gyromagnetic factor. 
As we see, this tensor $\hat \beta$ is not the mere transpose of the tensor $\hat \alpha$ given in \eqref{d-em-tensor} and, in contrast to $\hat \alpha$, does not contain information about magnon excitations ($\hat \chi^{{m},\pm \mathbf Q}$ does not enter here).

Let us now consider specific examples of magnetic structures. The first structure discovered with a long-period modulation was the helical one \cite{Izyumov84}:
\begin{align}
\mathbf M^{(0)} (\mathbf r)&=M_1 \cos ({\mathbf Q}\cdot {\mathbf r })\, \hat {\mathbf x} + M_3 \sin ({\mathbf Q}\cdot {\mathbf r })\,\hat {\mathbf z},
\label{helix}\end{align}
where $\mathbf Q = Q \, \hat {\mathbf y}$. This type of magnetic ordering is observed, for example, in CaFeO$_3$ \cite{Kawasaki98}. In this case, the inhomogeneous ME coupling \eqref{F_me} is ineffective in producing an electric polarization since this structure does not break inversion symmetry. Nevertheless, it gives rise to a dynamical ME effect. To the lowest order (i.e., considering the external field as the effective field acting on the magnetization in the Landau-Lifshitz equation), the non-zero components of the susceptibility $\hat \chi^{{m},\pm \mathbf Q}$ are 
\begin{subequations}\begin{align}
\chi^{{m},\pm \mathbf Q}_{xy}= -\chi^{{m},\pm \mathbf Q}_{yx} \propto M_z^{(0)}(\pm\mathbf Q),\\ 
\chi^{{m},\pm \mathbf Q}_{yz}= -\chi^{{m},\pm \mathbf Q}_{zy} \propto M_x^{(0)}(\pm\mathbf Q).
\label{}
\end{align}\end{subequations}
If the electric susceptibility $\hat \chi^{e}$ is diagonal, this means that the non-zero components of the dynamical ME tensors are 
$\alpha_{xx}$, $\alpha_{zz}$, $\beta_{xx}$ and $\beta_{zz}$. The ME response generated dynamically in this case turns out to be analogous to the static one of the prototypical magnetoelectric Cr$_2$O$_3$ (see, e.g., \cite{O'Dell,Krichevtsov93}).

Another important class of magnetic distributions is the cycloidal one: 
\begin{align}
\mathbf M^{(0)} (\mathbf r)&=M_2 \cos ({\mathbf Q}\cdot {\mathbf r })\, \hat {\mathbf y} + M_3 \sin ({\mathbf Q}\cdot {\mathbf r })\,\hat {\mathbf z},
\label{cycloidal}\end{align}
with $\mathbf Q = Q \, \hat {\mathbf y} $. The magnetization in $R$MnO$_3$ compounds, for example, develops this type of modulation, and its appearance is accompanied with ferroelectricity as we have explained above. The dynamic ME response in this case has the following features. The susceptibility $\hat \chi^{{m},\pm \mathbf Q}$ for the cycloidal has the non-zero components:
\begin{subequations}\begin{align}
\chi^{m,\pm \mathbf Q}_{xy}= -\chi^{m,\pm \mathbf Q}_{yx} \propto M_z^{(0)}(\pm\mathbf Q),\\ 
\chi^{m,\pm \mathbf Q}_{xz}= -\chi^{m,\pm \mathbf Q}_{zx} \propto M_y^{(0)}(\pm\mathbf Q),
\label{}
\end{align}\end{subequations}
to the lowest order. In consequence, for a diagonal electric susceptibility, the only nonzero component of the ME tensor $\hat \alpha$ turns out to be
\begin{align}
\alpha_{xy} (\mathbf q, \omega) = 
4i f Q M_2 \chi_{xx}^{e} (\mathbf q, \omega)\chi^{{m},-\mathbf Q}_{xy}(\mathbf q, \omega).
\label{alpha}
\end{align}
The fact that $\alpha_{xy} \not = \alpha _{yx}(=0)$ is a consequence of the inequivalence between the $x$ and $y$ directions in this magnetic structure. This reflects also in $\hat \beta$, whose non-zero component reduces to $\beta_{yx}$ as can be seen from \eqref{d-me-tensor}. 

The optical response of $R$MnO$_3$ compounds with this type of cycloidal magnetization shows absorption peaks at frequencies too small to be connected with pure phonon modes ($\sim \text{THz}$) \cite{Pimenov06,Sushkov07,Takahashi08,Kida08}. 
In TbMnO$_3$, in particular, these peaks have been correlated with the low-lying excitations of the cycloidal observed by inelastic neutron scattering experiments \cite{Senff07}. So they are interpreted as due to the electromagnon excitations naturally expected from the coupling \eqref{F_me} (see, e.g., \cite{Katsura07}). The electromagnon response to an external ac electric field is computed in \cite{Katsura07} as the (fluctuation) contribution to the electric permittivity due to the cycloidal excitations, and these results are further used to derive certain selection rules for the above optical experiments (see e.g. \cite{Takahashi08}). One has to realize that, however, this is not the whole story. Electromagnons actually react to both electric and magnetic components of the external radiation due to their hybrid character, so their final response can be more complex [Eqs. \eqref{d-em_coupling} and \eqref{d-me_coupling}, in general, do not reduce an effective permittivity]. 

Let us illustrate this point by computing the reflection coefficient for a vacuum-cycloidal magnet interface. For the sake of concreteness we restrict ourselves to the case of normal incidence and linear polarization along the principal axes of the magnet [which are assumed to be the axes of the cycloidal \eqref{cycloidal} in the following]. The result still depends on the orientation of the incident field with respect to the cycloidal. If the wavevector of the cycloidal $\mathbf Q$ is parallel to the interface the process is insensitive to the dynamical ME effect [$\hat \alpha$ and $\hat \beta$ do not enter the reflection coefficient, which is given by the standard Fresnel formula (see e.g. \cite{Landau-Lifshitz_ECM})]. The same happens if $\mathbf Q$ is perpendicular to the interface and the electric field is along the polar axis of the cycloidal. However, if the electric field is perpendicular to the polar axis (i.e., the incident fields are $\mathbf E^i \parallel \hat {\mathbf x}$ and $\mathbf H^i \parallel \hat {\mathbf z}$), the ME effect comes into play. In this case, the dispersion law for the light propagating through the magnet is
\begin{align}
ck = \pm {\sqrt{\Big(\varepsilon - {\alpha \beta \over \mu}\Big)\mu} }\, \omega, 
\label{}
\end{align}
and the reflection coefficient is found to be
\begin{align}
r = {1 - \sqrt{\Big(\varepsilon - {\alpha \beta \over \mu}\Big){1 \over \mu}}
\over 
1 + \sqrt{\Big(\varepsilon - {\alpha \beta \over \mu}\Big){1 \over \mu}}}
\label{}
\end{align}
(hereafter we drop subindices since we are dealing with the only non-zero components of the ME tensors). In these expressions we see that the dynamic ME effect results in a new (effective) permittivity $\varepsilon_\text{eff} = \varepsilon - {\alpha \beta \over \mu}$ that now has poles at the magnon frequencies because of $\alpha \sim \chi ^{m,\mathbf Q}$ [see Eq. \eqref{alpha}], i.e., this effective permittivity has electromagnon features. This is in tune with \cite{Katsura07} and the general interpretation of the experimental data (see e.g. \cite{Kida08}). For other orientations, however, the actual situation turns out to be a bit more subtle. 

If, for example, the plane of the cycloidal is parallel to the interface and the electric field is directed along the polar axis ($\mathbf E^i \parallel \hat {\mathbf z}$ and $\mathbf H^i \parallel \hat {\mathbf y}$), the ME coupling effectively result in the modification of the magnetic permeability (not the electric permittivity as before). This is not captured in \cite{Katsura07} because only the influence of the electric field is taken into account. Experimentally, however, no electromagnon feature seems to be observed for this orientation \cite{Takahashi08}. 
\textcolor{black}{
The reason may be the fact that the magnetolelectric effect is here reduced just to an effective permeability (since the magnetic response alone is generally quite weak) and/or the necessity of going beyond the isotropic coupling \eqref{F_me} for these materials.}
Furthermore, if the electric field is perpendicular to the plane of the cycloidal and this plane is perpendicular to the interface ($\mathbf E^i \parallel \hat {\mathbf x}$ and $\mathbf H^i \parallel \hat {\mathbf y}$), the dispersion law is obtained from the equation: 
\begin{align}
\Big({ck\over \omega} - \alpha\Big)\Big({ck\over \omega} - \beta\Big) = 
\varepsilon\mu.
\label{me-dispersion}
\end{align}
In this case the ME effect plays a genuine role, not reducible to a mere modification of the electric permittivity (or magnetic permeability) as before. The waves associated with the two solutions of \eqref{me-dispersion}, for example, have different phase velocities. This possibility of removing the degeneracy between forward and backward waves is known long ago in genuine magnetoelectrics \cite{O'Dell,Arima08}. 

It is worth mentioning that the field transmitted into the magnet can acquire a longitudinal component due to the ME effect. In the case $\mathbf E^i \parallel \hat {\mathbf x}$ and $\mathbf H^i \parallel \hat {\mathbf z}$, for example, the transmitted field is such that
\begin{align}
{H_y^{t} \over H_z^{t}} = -{\beta \over \big(\varepsilon - {\alpha \beta\over \mu}\big){1\over \mu}}.
\end{align}
This possibility is also known for ordinary magnetoelectrics. To probe experimentally this longitudinal component can be somewhat difficult, but there is a related aspect of the ME effect whose experimental verification is, at least conceptually, much easier. It is the possibility of having a ME rotation of the reflected light. This possibility is quite obvious for the helical structure \eqref{helix} taking into account that its ME response is analogous to that of Cr$_2$O$_3$ as we have seen. A similar rotation  (see, e.g., \cite{Krichevtsov93}) is therefore expected, with the particularity that in helical magnets like CaFeO$_3$ it can be significantly enhanced due to the resonant behavior of the ME response. 
\textcolor{black}{
This dependence on the frequency is absent in conventional magnetoelectrics \cite{Hornreich68}. In spiral magnets, it is closely related to the electromagnon features in their spectrum
\cite{estimates-rot}.}

Let us now explore yet another phenomenon that might benefit from these features.
The ME effect has been pointed out as an interesting possibility to achieve a negative refractive index behavior \cite{Pendry04,Tretyakov05}, recently demonstrated experimentally \cite{Zhang09}. The key point in these experiments is the fabrication of metamaterials with chiral constituents. As in ordinary negative index metamaterials, the achievement of a negative index regime relies on the resonant response of the resulting system. In these cases, one basically deals with the resonances of the constituent particles \cite{Tretyakov05}. This imposes severe limits to the range of frequencies at which the corresponding negative index behavior can be achieved. In the case of spiral magnets, on the contrary, it is the collective behavior of the system what gives rise to the ME effect. But the resulting resonant behavior [see Eqs. \eqref{d-em-tensor} and \eqref{d-me-tensor}] is basically the same than in chiral metamaterials \cite{Tretyakov05}. Consequently these type of magnetic structures may also result in a negative index behavior, now at the frequencies of the corresponding electromagnons ($\sim$ THz for natural compounds).

It is worth mentioning that spatial dispersion effects \cite{Landau-Lifshitz_ECM,Agranovich} can also be dynamically amplified in spiral magnets. 
Generally spatial dispersion produces minute effects in optical experiments and, in practice, the response of the system is well described by the limiting $q\to 0$ behavior of the electric and magnetic susceptibilities $\hat \chi^{e(m)}$ \cite{Landau-Lifshitz_ECM,Agranovich}. 
Accordingly, in the computation of the ME response tensors $\hat \alpha $ and $\hat \beta$ we have neglected terms $\mathcal O (q)$ coming from the inhomogeneous ME coupling \eqref{F_me} [see Eq. \eqref{motion_p}]. 
Close to the electromagnon resonances, however, this neglection might be unjustified since these terms are dynamically enhanced by the same resonant mechanism that operates for $\hat \alpha (q =0)$ and $\hat \beta (q =0)$. The inhomogeneous ME coupling then has to be considered in its full extent \cite{note}, and spatial dispersion effects may compete with the dynamic $q=0$ ME response in a similar way that it does, for example, with the static ME effect in Cr$_2$O$_3$ \cite{Krichevtsov93}.

In summary, we have shown that the optical response of spiral magnets exhibits electromagnon features encoded in the form of a resonant magnetoelectric response. Spiral ordering does not have to be accompanied with multiferroicity (and/or a static magnetoelectric effect) to have these features. We have discussed the role of this dynamical response in optical experiments on multiferroics, showing that the observed electromagnon features not always can be reduced to an effective electric permittivity. We also have argued that electromagnon resonances in spiral magnets can amplify spatial dispersion and nonreciprocal effects. These resonances may also permit to achieve a negative refractive index behavior.

I acknowledge P. Bruno, A. Levanyuk and especially E. Kats for very fruitful discussions.


\begin{references}
\bibitem{note} The most general form of the (isotropic) inhomogeneous ME coupling is $- f_1 {\mathbf P} \cdot {\mathbf M}(\nabla \cdot {\mathbf M}) 
- f_2 {\mathbf P} \cdot [{\mathbf M}\times (\nabla \times {\mathbf M})]$. For the dynamical effect we consider, only its antisymmetric part \eqref{F_me} is relevant in the long-wavelength limit [$f= f_1+f_2$ in Eq. \eqref{F_me}]. This is not the case, however, close to wavevectors of the magnetic modulation and/or if we are interested in spatial dispersion effects, to which the two terms above contribute separately \cite{Cano08}.
\bibitem{estimates-rot} The frequency window in which the rotation is expected to vary with the frequency is determined by the electromagnon damping. The variation can be $\pi /4$ for very sharp electromagnon peaks. For a ME response like that observed in spiral multiferroics rotations of $\sim \pi/40$ can be expected within a frequency window of $\sim 10 \, \text{cm}^{-1} $ about the corresponding electromagnon peaks.
\end{references}
\end{document}